\documentclass[10pt,a4paper]{article}
\usepackage{amsmath}
\usepackage{amsfonts}
\usepackage{amssymb}
\usepackage[dvips,letterpaper,margin=0.75in,bottom=0.5in]{geometry}
\usepackage[T1]{fontenc}
\usepackage[utf8]{inputenc}
\usepackage{authblk}
\usepackage{scrextend}
\usepackage{footmisc}
\usepackage{indentfirst}
\usepackage{graphicx} 
\graphicspath{{figures/}}
\graphicspath{ {images/} }
\usepackage{epstopdf}
\usepackage{pgfplots}
\usetikzlibrary{calc}

\title{Sections and Chapters}
\usepackage{cite}
\title{A Perspicuous Description of the Schwarzschild Black Hole Geodesics}
\author{Metin Arik$^{1,}$\footnote{metin.arik@boun.edu.tr}}
\author{M. Tuna Pesen$^{1,}$\footnote{tuna.pesen@boun.edu.tr}}
\affil{$^{1}$ Bogazici University, Department of Physics, Istanbul, Turkiye}

\begin{document}

\maketitle
\begin{abstract}
Schwarzschild black hole is the simplest black hole that is studied most in detail. Its behavior is best understood by looking at the geodesics of the particles under the influence of its gravitational field. In this paper, the focus of attention is giving a perspicuous description of the Schwarzschild geodesics by using analogue potential approach. Specifically we discuss geodesics of light and of a massive particle in the case that their angular momentum is non zero in the Schwarzschild spacetime. This discussion is done by defining analogue potentials out of geodesic equations and defining relevant dimensionless conserved quantities. Then, we designate how geodesics change in response to the change of these quantities. Our results indicate the relation between the particles' motion near black hole horizon and their angular momentum. Furthermore, we make a comparison between Newtonian Physics (NP) and General Relativity (GR) in the language of the analogue potential approach.
\end{abstract}
\section{Introduction}
Although Schwarzschild black hole is the simplest black hole there are still discussions associated with it. Especially the questions, raised by the coordinate singularity at the event horizon, about the interior of the black hole are still debatable \cite{Doran:2006dq,Francis:2003rj}. In literature it is generally accepted that beyond of the event horizon physical world losses its meaning. All the particles that pass the horizon end at the singularity centered at the black hole.  On the other hand, there are some studies which point that the spacetime beyond of the event horizon can be described in Kruskal-Szekeres coordinates  \cite{Araya:2015fva} . To get a better understanding of the black holes, geodesic calculations play a crucial role. In this context, geodesics in Schwarzschild spacetime can be completely solved analytically using Weierstrass and Jacobi elliptic functions\cite{Hackmann:2015ewa,Gibbons:2011rh}. It is also common to study geodesics of the particles through an analogue potential approach \cite{Ghosh:2015eya,Grib:2014afa}. This approach clearly reveals Newtonian results for large $r$ limit and can be used for modified Schwarzschild geodesics too \cite{Halilsoy:zva}. Our motivation on this paper is to formalize Schwarzschild black hole geodesics by using dimensionless parameters related to the angular momentum of light and a massive particle so that the energy equations become simple. The equations for the geodesics can be solved for the four cases: for light and for a massive particle, $m$, in the absence or the presence of the angular momentum. We consider the two cases for which the angular momentum is non zero for light and the mass $m$. By starting from the metric of each case we solve the equations for the geodesics. Then out of them we form the energy equations  and define analogue potentials associated with each case. After this we investigate the behavior of the geodesics in response to the change in angular momentum. For light it is common to define an inverse radius, $u=1/r$, \cite{Halilsoy:zva} like in the  Kepler problem, but we leave it as $1/r$ throughout the calculations. 
\section{Introducing the analogue potential concept}
According to NP the attractive force, $\vec{F}$, exerted on a body with mass $m$, by another mass $M$ (where $M$ and $m$ are assumed to be spherical or point masses) located at the origin is given by
\begin{equation}
\vec{F}=-\frac{GMm}{r^2}\hat{r}
\end{equation}
where $G$ is the universal gravitation constant and $r$ is the distance from origin to $m$. 
Thus, the gravitational potential, $\Phi$,  generated by $M$ satisfies \textit{Poisson's Equation} 
\begin{equation}
\nabla ^2 \Phi =4\pi GM\delta(\vec{r})
\end{equation}
where $M\delta(\vec{r})  $ is the mass density obtained by taking $M$ as a point particle at the origin. The spherically symmetric solution of above equation gives the gravitational force as
\begin{equation}
\begin{split}
\vec{F} &= -m\vec{\nabla}\Phi \\
\vec{g}   &= -\vec{\nabla}\Phi
\end{split}
\end{equation}
where $\vec{g}$ is the gravitational field. The total energy, $E$, of the particle, $m$, is a conserved quantity and is equal to sum the of its kinetic and potential energies
\begin{equation}
\label{n4}
E = \frac{1}{2}mv^2 + \Phi = const
\end{equation}
With this equation it is easy to keep track of particles, i.e. the equations of motion. However in GR, the attraction between particles (both massive and massless) is stated to be a consequence of the curvature of the spacetime or simply the "geometry". Hence in the GR regime one can not refer to "force" or "potential" like in NP. In GR, gravity is described by \textit{Einstein Field Equations}
\begin{equation}
G_{\mu\nu}=\frac{8\pi G}{c^4}T_{\mu\nu}
\end{equation}
where $ G_{\mu\nu} $ is the \textit{Einstein Tensor}  and $T_{\mu\nu}$ is the \textit{Stress-Energy Tensor}.
Einstein equations describe the geometry of spacetime based on the matter content in that spacetime and motion of the matter is determined by this geometry. Describing the evolution of particles needs first writing the metric, $ds^2$, which contains the information of geometry of the spacetime 
\begin{equation}
\label{metric}
ds^2=g_{\mu\nu}dx^\mu dx^\nu
\end{equation}
This metric description is actually the \textit{Pythagorean Theorem} that is not only valid for flat but also for curved spacetimes. 
Spherically symmetric vacuum solution of the Einstein field equations gives the Schwarzschild solution of which its metric is given by
\begin{equation}
\label{schw}
ds^2 = \left(1-\frac{2GM}{c^2r}\right)c^2dt^2 - \left(1-\frac{2GM}{c^2r}\right)^{-1}dr^2 - r^2d\theta^2  -r^2\sin^2\theta d\phi^2 
\end{equation}
Solving the Euler-Lagrange equations by defining an analogue Lagrangian from the geodesic action reveals equations for the evolution of the particle. As can be seen in the next sections, the terms in the equations derived from the metric are very similar to the potential terms in NP. Therefore it is useful to define such terms as \textit{analogue potentials}. This approach is worthwhile by means of seeing the difference between NG and GR too.

\section{Non-null geodesic with non-zero angular momentum}
In the Schwarzschild spacetime a non-null metric is written as
\begin{equation}
\label{1}
ds^2 = \left(1-\frac{2GM}{c^2r}\right)c^2dt^2 - \left(1-\frac{2GM}{c^2r}\right)^{-1}dr^2 - r^2d\phi^2
\end{equation}
where we set $\theta= \pi/2$ without loss of generality and $(r,\theta, \phi)$ are the usual spherical coordinates. The geodesic equations of the particle with unit mass are found by varying the action, $S$
\begin{equation}
\label{1a}
S=\int \sqrt{ds^2}=\int \mathcal{L}d\sigma
\end{equation}
where $\mathcal{L}$ is the Lagrangian and $\sigma$ is the arbitrary parametrization variable. Since $S$ is reparametrization invariant   \cite{Deriglazov:2011as} we can choose $d\sigma=ds$. This choice further reveals 
\begin{equation}
\label{1aa}
\mathcal{L}=\sqrt{c^2\left(1-\frac{2GM}{c^2r}\right)\left(\frac{dt}{ds}\right)^2 - \left(1-\frac{2GM}{c^2r}\right)^{-1}\left(\frac{dr}{ds}\right)^2 - r^2\left(\frac{d\phi}{ds}\right)^2} =1
\end{equation} 
Then solving the Euler-Lagrange equations gives the geodesic equations of the particle 
\begin{equation}
\left(1-\frac{r_s}{r}\right)c^2\frac{dt}{ds} = \frac{\epsilon}{c} 
\end{equation}
\begin{equation}
\frac{d\phi}{ds}=\frac{l}{r^2c} 
\end{equation}
\begin{equation}
\left(\frac{dr}{ds}\right)^2 =   \frac{\epsilon^2}{c^4} + \left(\frac{r_s-r}{r^3}\right)\frac{l^2}{c^2} + \left(\frac{r_s-r}{r}\right)
\end{equation}
where  $r_s\equiv \frac{2GM}{c^2}$, $\epsilon$ is the energy per mass  and $l$ is the angular momentum per mass of the particle. Equation (13) is found by dividing both sides of the metric by $ds^2$ and using (12) and (11). We write (13) in a way that it resembles the energy equation of a non relativistic particle by defining an analogue potential, $V_{a}(r)$,
\begin{equation}
\left(\frac{dr}{ds}\right)^2 + V_{a}(r) = E
\end{equation}
where \textit{E} and  $V_{a}(r)$ are defined to be
\begin{equation}
E=\frac{\epsilon^2}{c^4}-1
\end{equation}
\begin{equation}
V_a(r)= - \frac{1}{x} + \frac{\lambda^2}{x^2} - \frac{\lambda^2}{x^3}
\end{equation}
where we define a dimensionless conserved constant as $\lambda  \equiv  l/r_sc$ 
and the variable $r$ is scaled as $x \equiv r/r_s$. For a few representative $\lambda^2$ values, the graph $V_a(r)$ vs. $x$ can be seen in FIG.1.
\begin{figure}[h]
\centering
\includegraphics[width=3in]{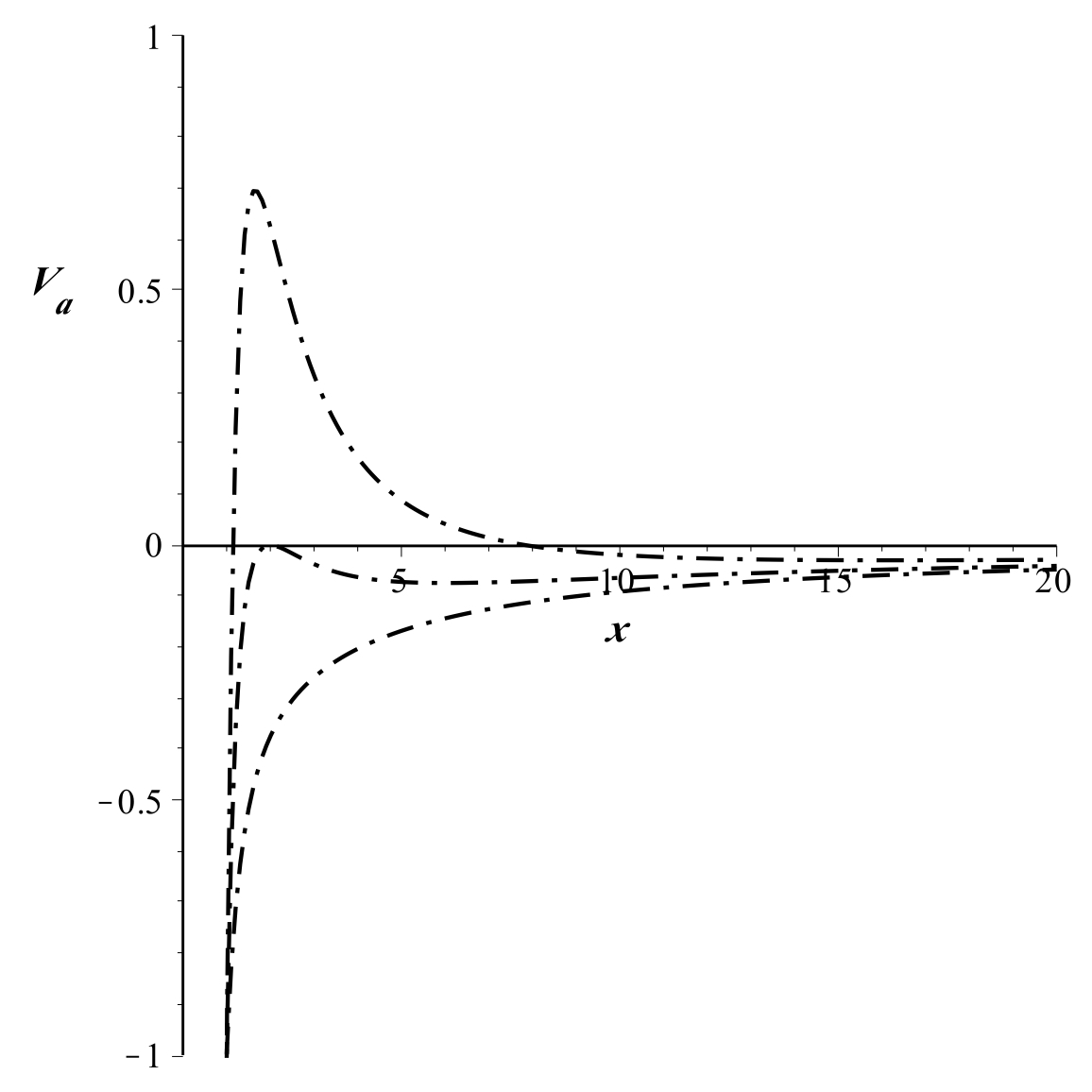}
\caption{Effective potential for $m$, $V_{a}(r)$, as a function of  $x\equiv r/r_s$ for a few representative values of $\lambda^2  \equiv  \left(\frac{l}{r_sc}\right)^2$ where curves corresponds to $\lambda^2$: (1, 4, 9) from bottom to top.}
\end{figure}
As shown in the figure, for $\lambda^2=1$, the particle with $E=0$ coming from infinity falls into the center of the potential. For $\lambda^2  = 4$, the particle with $E = 0$ coming from infinity, there is an unstable circular orbit at $r = 2r_s$ so that it may fall in or escape back to infinity rather than orbiting circularly forever. In addition to this, there is a stable circular orbit at the minimum of the potential, which corresponds to $r = 6r_s$. For $\lambda^2=9$, there exist one unstable orbit, one stable orbit and infinite number of precessing orbits for the particle. The particle, on the other hand, coming from infinity with the energy greater than the maximum value of $V_a(r)$, falls into the center of the potential.
\section{Null geodesic with non-zero angular momentum}
In the Schwarzschild spacetime a null metric is given by
\begin{equation}
ds^2=\left(1-\frac{2GM}{c^2r}\right)c^2dt^2-\left(1-\frac{2GM}{c^2r}\right)^{-1}dr^2-r^2d\phi^2=0
\end{equation}
where we take $\theta=\pi/2$ without loss of generality. The equations for the geodesic of light are obtained by solving Euler-Lagrange equations. We parametrize the Lagrangian, $\mathcal{L}$, with an arbitrary parameter $\sigma$
\begin{equation}
S = \int \sqrt{ds^2} =\int L d\sigma
\end{equation}
Equations for the geodesics are
\begin{equation}
\left(1-\frac{r_s}{r}\right)c\frac{dt}{d\sigma} = \frac{\epsilon}{c^2} ,
\end{equation}
\begin{equation}
\frac{d\phi}{d\sigma}=\frac{l}{r^2c}
\end{equation}
\begin{equation}
\left(\frac{dr}{d\phi}\right)^2 - \left(\frac{\epsilon}{lc}\right)^2r^4 + r^2 - r_s r  = 0
\end{equation}
where (21) is found using Hamilton-Jacobi method and eliminating $d\sigma$, using (19) and (20). We shall write (21) by defining an analogue potential, $V_a(r)$. In this way the analogue energy equation for light can be constructed as 
\begin{equation}
\label{6a}
\left(\frac{dr}{d\phi}\right)^2 + V_a(r)=0
\end{equation}
where $\phi$ plays the role of time, $t$. We write $V_a(r)$ as a dimensionless equation so that a graph with dimensionless variables can be plotted
\begin{equation}
\label{7a}
\frac{V_a(r)}{r_s^2} = -\left(\frac{r_s}{r_l}\right)^2 x^4 +x^2-x
\end{equation}
where $x\equiv r/r_s$ and $r_l \equiv l c / \epsilon$. The graph of $V_a(r)/r^2_s$ for a few representative $\left(\frac{r_s}{r_l}\right)^2$ values can be seen on FIG.2.
\begin{figure}[h]
\centering
\includegraphics[width=3in]{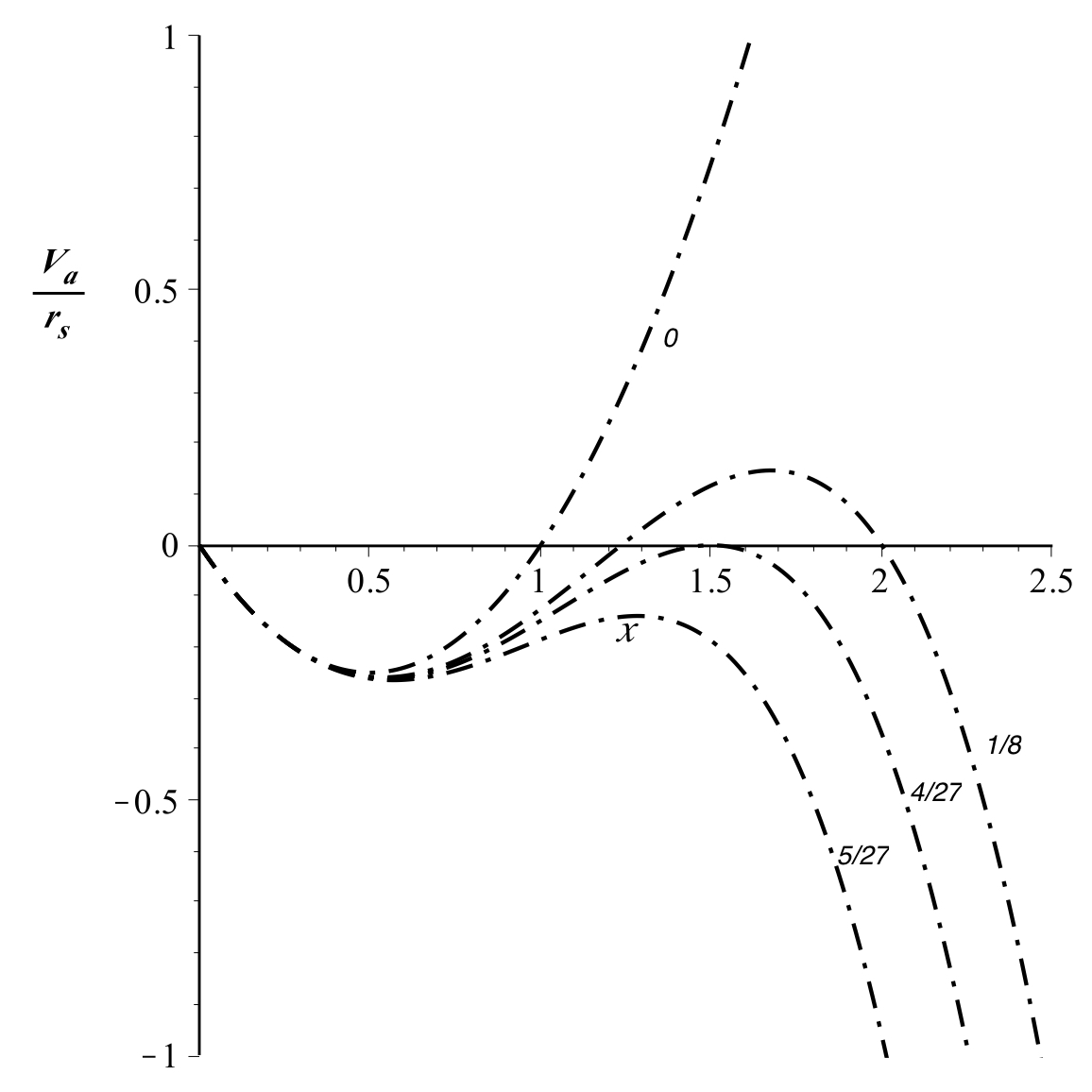}
\caption{Normalized analogue potential for light, $V(r)/r_s$, as a function of  $x\equiv r/r_s$ for a few representative values of $\left(\frac{r_s}{r_l}\right)^2$, where curves correspond to $\left(\frac{r_s}{r_l}\right)^2: (0,\frac{1}{8}, \frac{4}{27}, \frac{5}{27})$ from top to bottom.}
\label{four}
\end{figure}
As follows from the figure, null geodesics stay in the horizon only when $\left(\frac{r_s}{r_l}\right)^2 \rightarrow 0$ which corresponds to the $l \rightarrow 0$ limit. However for the other cases, that is $r_l$ is finite, null geodesics seem to reach beyond the horizon ($x=1$). Furthermore, with respect to the curve with $\left(\frac{r_s}{r_l}\right)^2=5/27$, the null geodesic seems to extends to infinity. 
\section{Comparison between GR and NP} 
We write the analogue potential for non-null geodesics defined in (16) in a clearer way
\begin{equation}
V_{a}(r)= -\frac{r_s}{r} + \frac{l^2}{c^2r^2} - \frac{r_sl^2}{c^2r^3}
\end{equation}
Noticing that first two terms are the same as gravitational and centrifugal potential terms in Newtonian effective potential, $V_N$. Again with the substitution of $x\equiv r/r_s$ and $\lambda \equiv l/r_sc$ we have
\begin{equation}
\label{52}
V_a(r)= \underbrace{- \frac{1}{x} + \frac{\lambda^2}{x^2}}_{\text{$V_{N}$}} -\frac{\lambda^2}{x^3}
\end{equation}
\begin{equation}
\label{522}
V_N=- \frac{1}{x} + \frac{\lambda^2}{x^2} 
\end{equation}
An important implication here is that the presence of $-\lambda^2/x^3$ in $V_a$ which is absent in $V_N$.  Actually this term is the correction coming from general relativity. In (25), for large $r$, first two terms are important and we can neglect the third one but for small $r$, the last term can not be negligible anymore. Also paying attention  that the correction term comes with the minus sign, which shows that it contributes to the potential in an attractive way. Furthermore, there is a finite centrifugal potential barrier in GR whereas it is infinite in NP. Clearly, addition of the correction term turns the infinite potential barrier into a finite barrier. To see the effect of addition of this correction we draw the graph of those two potentials, $V_a(r)$ and  $V_N$, for $\lambda^2=10$.
\begin{figure}[h]
\centering
\includegraphics[width=3in]{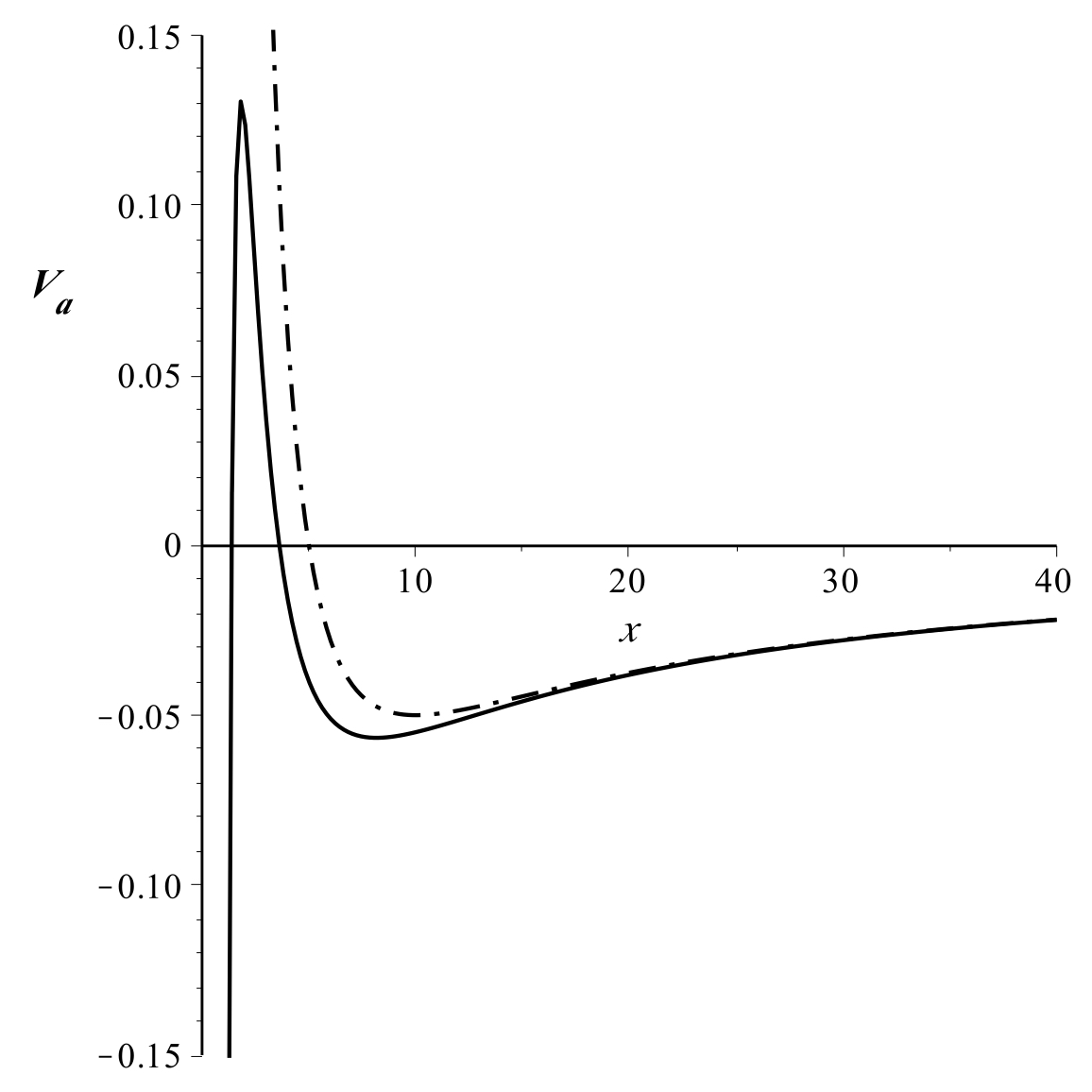}
\caption{Relativistic correction to the Newtonian effective potential. The dashed and solid curves shows the $V_N$ and $V_a$ respectively. The curves become identical for $r>>r_s$.}
\label{comparison}
\end{figure}
According to FIG.3 it is obvious that, a particle with energy $E$ can never pass the Newtonian potential hence never plunges into the center of the potential, whereas maximum point of the solid line shows that there is a finite potential barrier for the particle near the Schwarzschild black hole. So, whenever it has a greater energy than the maximum of  $V_{a}(r)$, the particle plunges into the black hole. FIG.3 also indicates that for large $r$ both potentials have the same asymptotic behavior as expected which is interpreted as NP being a large $r$ approximation of GR.
\section{Conclusion}
We have shown that the analogue potential approach is able to give a clear picture of how orbits around the Schwarzschild black hole behave. In our treatment, for massive particles, $s$ plays the same role which $t$ plays in non-relativistic physics. On the other hand for null geodesics we have noticed that analogue energy equation for the normal variable of time, $t$, is complicated. Therefore we have used angular variable $\phi$ to describe the evolution of the geodesic of light. By doing this we get a simple energy equation for light and draw the analogue potential for a few representative values of angular momentum. Treating $\phi$ as $t$ reveals some interesting results;
according to our graph, light with a finite amount of angular momentum is able to reach beyond the event horizon whereas, light stays in the horizon only when its  angular momentum goes to infinity. 
\bibliographystyle{unsrt}
\bibliography{arxivref}
\end{document}